\documentclass[aps,prl,reprint,superscriptaddress,showpacs]{revtex4-1}

\usepackage{amssymb,amsmath}
\usepackage{graphicx}
\usepackage{hyperref}

\begin{document}

\title{Spin multistability of cavity polaritons in a magnetic field}

\author{S. S. Gavrilov} \email{gavr\_ss@issp.ac.ru}
\affiliation{Institute of Solid State Physics RAS, Chernogolovka,
  142432, Russia}

\author{A. V. Sekretenko}
\affiliation{Institute of Solid State Physics RAS, Chernogolovka,
  142432, Russia}

\author{N. A. Gippius}
\affiliation{A.\,M.\,Prokhorov General Physics Institute, RAS, Moscow
  119991, Russia}
\affiliation{LASMEA, UMR 6602 CNRS, Universit\'{e} Blaise Pascal,
  63177 Aubi\`{e}re, France}

\author{C. Schneider}
\affiliation{Technische Physik, Physikalisches Institut and Wilhelm
  Conrad R\"ontgen Research Center for Complex Material Systems,\\
  Universit\"at W\"urzburg, D-97074 W\"urzburg, Germany}

\author{S. H\"ofling}
\affiliation{Technische Physik, Physikalisches Institut and Wilhelm
  Conrad R\"ontgen Research Center for Complex Material Systems,\\
  Universit\"at W\"urzburg, D-97074 W\"urzburg, Germany}

\author{M. Kamp}
\affiliation{Technische Physik, Physikalisches Institut and Wilhelm
  Conrad R\"ontgen Research Center for Complex Material Systems,\\
  Universit\"at W\"urzburg, D-97074 W\"urzburg, Germany}

\author{A. Forchel}
\affiliation{Technische Physik, Physikalisches Institut and Wilhelm
  Conrad R\"ontgen Research Center for Complex Material Systems,\\
  Universit\"at W\"urzburg, D-97074 W\"urzburg, Germany}

\author{V. D. Kulakovskii}
\affiliation{Institute of Solid State Physics RAS, Chernogolovka,
  142432, Russia}

\date{\today}

\begin{abstract}
  Spin transitions are studied theoretically and experimentally in a
  resonantly excited system of cavity polaritons in a magnetic field.
  Weak pair interactions in this boson system make possible fast and
  massive spin flips occurring at critical amplitudes due to the
  interplay between amplitude dependent shifts of eigenstates and the
  Zeeman splitting.  Dominant spin of a condensate can be toggled
  forth and back by tuning of the pump intensity only, which opens the
  way for ultra-fast spin switchings of polariton condensates on a
  picosecond timescale.
\end{abstract}


\maketitle

Collective phenomena in cavity-polariton systems attract a broad
interest during the last decade.  Being weakly interacting bosons
formed due to strong exciton-photon coupling in
microcavities~\cite{Weisbuch92}, polaritons exhibit collective
properties typical of both matter waves and light waves, including
optic parametric scattering~\cite{Stevenson00} and Bose
condensation~\cite{Kasprzak06}.
The combination of non-zero spin, spin-sensitive interactions and the
ability of polaritons to form macro-occupied states allows studying a
row of new spin phenomena manifesting themselves in the transitions
between distinct spin states of multi-component condensates.
Cavity-polariton systems are much more convenient to explore by
optical means than condensates in He-3 or other Bose systems with
non-zero spin those, however, may exhibit similar collective
properties due to spin-dependent interactions.

Under resonant excitation, polariton-polariton interactions lead to a
multistable behavior of polariton condensates with several steady
states feasible at fixed pump
parameters~\cite{Gippius07,Shelykh08-prl,Liew08-prl-neur,
  Gavrilov10-en,Paraiso10,Sarkar10,Adrados10,Gavrilov12-prb,
  Gavrilov12-prb-strains,Bozat12,Gavrilov13-apl}.  A switching between
the steady-state branches nearby bifurcation points is accompanied by
sharp jumps in the cavity-field intensity and polarization.  The state
of the spin-up or spin-down component of the condensate is switched
due to a feedback loop between its amplitude and effective resonance
frequency.  The same kind of transitions is also responsible for
many-mode optical parametric
oscillation~\cite{Krizhanovskii08,Demenev08}, spin ring
patterns~\cite{Shelykh08-prl,Sarkar10,Adrados10,Gavrilov12-prb}, and
bright polariton solitons~\cite{Sich11}, which all would be impossible
without polariton bi- or multistability.  The energy-shifting
interactions are also known to cause spin transitions in
non-resonantly pumped polariton condensates in a magnetic
field~\cite{Larionov10,Walker11}.

By now, many-branch multistability of polariton condensates has been
evidenced under continuous-wave (cw) excitation in few $\mu$m wide
cavity mesas~\cite{Paraiso10}.  It was shown that the average spin of
the condensate can be triggered from almost $-1$ to $+1$ by varying
the degree of circular polarization of the pump wave within a
comparatively narrow interval.  However, changing the pump intensity
at a fixed pump polarization cannot alter the dominant spin of the
condensate that, in this case, is pinned to dominantly right or left
polarization of the pump wave.  The necessity to tune the input
polarization lefts open the question on how fast such spin inversions
of condensates can be, since a controllable tuning of light
polarization on a picosecond timescale (that is comparable to the
lifetime of cavity polaritons) is quite problematic.

In this Letter we study transitions between distinct spin states of
many-polariton systems resonantly excited in a planar cavity placed
into a magnetic field.  The condensates are shown to have coexistent
opposite-spin states in a wide range of pump powers and polarizations.
We found that the interplay between blue-shifts of eigenstates and the
Zeeman effect releases the spin pinning, so that a condensate can be
switched between opposite-spin states by varying the intensity even
keeping fixed the polarization of the pump wave.  As such, it
constitutes a qualitatively new kind of multistability that enables
spin-inverting transitions of multi-component bosonic condensates
within extremely short time intervals under pulsed photoexcitation.
We have tracked the cavity transmission under 100 ps pump pulses and
found the time of the transition between opposite-spin states of a
polariton condensate to be about ten picoseconds.

\textit{Multistability and steady-state solutions.}---Resonantly
driven cavity-polariton condensates are usually considered within a
coherent mean-field approach based on the Gross-Pitaevskii equations.
For simplicity we write them for the single lower polariton mode
driven by a coherent pump with zero in-plane wave vector ($\mathbf k =
0$)~\cite{Gippius07,Shelykh08-prl,Liew08-prl-neur,Gavrilov10-en,Walker11}:
\begin{multline}
  \label{eq:gpe}
  i \hbar \frac{d}{dt} \binom{\psi_+}{\psi_-} = \left( \hat E_0 - i
    \hat \gamma_0 \right) \binom{\psi_+}{\psi_-} \\ + \binom {(V_1
    |\psi_+|^2 + V_2 |\psi_-|^2) \psi_+} {(V_2 |\psi_+|^2 + V_1
    |\psi_-|^2) \psi_-} + \dbinom{F_+(t)}{F_-(t)}.
\end{multline}
Herein are $ F_\pm (t) = \sqrt{\frac{I_p(t)}{2} (1 \pm \rho_p)} e^{i
  (E_p / \hbar) t + i \phi_\pm}$; $I_p = |F_+|^2 + |F_-|^2$ the pump
intensity, $\rho_p$ the degree of circular polarization, $E_p$ the
pump energy, and $\Phi_p = \phi_+ - \phi_-$ the phase shift between
the $\sigma^\pm$ components of the pump wave.  $V_{1, 2}$ are the
polariton-polariton interaction constants.  The polariton mode is
characterized by its energy $\hat E_0$ and decay rate $\hat \gamma_0$
which are $2 \times 2$ matrices written in the $\sigma^\pm$ basis.
For simplicity we assume $\hat \gamma_0$ to be spin-symmetric, $\hat
\gamma_0 = \gamma_0 \hat\openone$, whereas $\hat E_0$ has the form
\begin{equation}
  \label{eq:2}
  \hat E_0 = E_0 \hat\openone + \frac{\delta_c}{2}
  \begin{pmatrix}
    1 & 0 \\ 0 & -1
  \end{pmatrix}
  + \frac{\delta_l}{2}
  \begin{pmatrix}
    0 & 1 \\ 1 & 0
  \end{pmatrix}.
\end{equation}
Its first term is the energy of polaritons in a symmetric structure,
the second term provides the $\sigma^\pm$ splitting, $E_0^+ - E_0^- =
\delta_c$, and the third term gives the X-Y splitting $E_0^{(x)} -
E_0^{(y)} = \delta_l$ in the Cartesian basis
$
  \binom{\psi_x}{\psi_y} = \frac{1}{\sqrt 2}
  \bigl( \begin{smallmatrix}
    1 & 1 \\ i & -i
  \end{smallmatrix} \bigr)
  \binom{\psi_+}{\psi_-}.
$
The X-Y splitting 
may be due to crystalline disorder
or TE-TM splitting of cavity photons at $\mathbf k \neq 0$.

The interaction between polaritons with anti-parallel spins ($V_2$) is
attractive and typically much weaker than that between polaritons with
parallel spins ($V_1$)~\cite{Renucci05,Kavokin.K05,Vladimirova10};
below we assume $V_2 = -0.1 V_1$.  The units for $\psi$ can be fixed
by the condition $V_1 = 1$, so that $|\psi|^2$ has the dimension of
energy and, in the case of circularly polarized excitation, coincides
with the blue-shift of the driven mode.  In the following simulations
we set $\gamma_0 = 0.05$\,meV and $\Delta = E_p - E_0 = 0.2$~meV.

By substituting $\psi_\pm(t) = \bar \psi_\pm e^{-i (E_p / \hbar) t}$
into Eq.~(\ref{eq:gpe}) one gets a time-independent equation that
allows one to find the response of the driven condensate mode on a
constant pump, $\bar\psi_\pm = \bar\psi_\pm(I_p, \rho_p, \Phi_p)$.
The problem concerned with asymptotic stability of the steady-state
solutions was considered in Ref.~\cite{Gavrilov10-en}.

\begin{figure}
  \centering
  \includegraphics[width=0.90\linewidth]{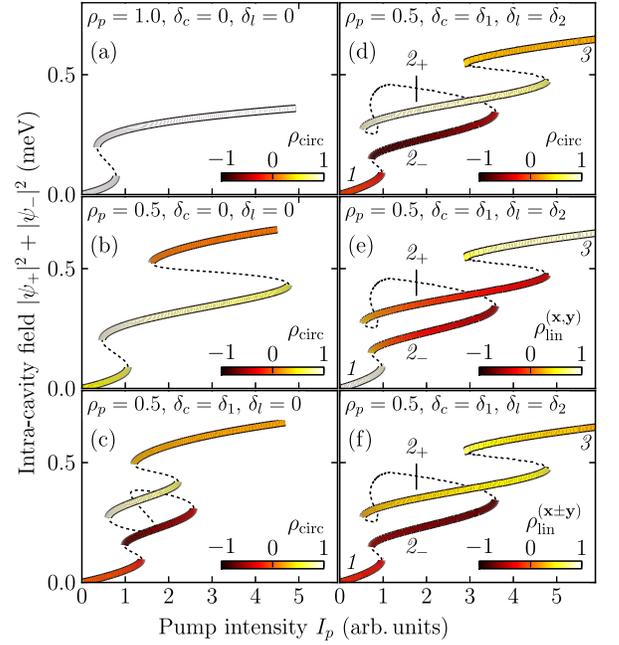}
  \caption{(Color online) The steady-state response of the driven
    polariton mode, $|\psi_+|^2 + |\psi_-|^2$ vs.\ $I_p$, for several
    combinations of $\rho_p$, $\delta_c$, and $\delta_l$ which are
    indicated at the top of each panel ($\delta_1 = -0.12$~meV,
    $\delta_2 = 0.05$~meV, $\Phi_p=0$).  The branches of
    asymptotically stable solutions are shown by colored heavy lines,
    the color indicates the degrees of circular polarization (a--d)
    and linear polarization in $(\mathbf x, \mathbf y)$ basis (e) and
    $(\mathbf x + \mathbf y, \mathbf x - \mathbf y)$ basis (f);
    unstable branches are shown by dashed lines.}
  \label{fig:fig1}
\end{figure}
\begin{figure}
  \centering
  \includegraphics[width=0.85\linewidth]{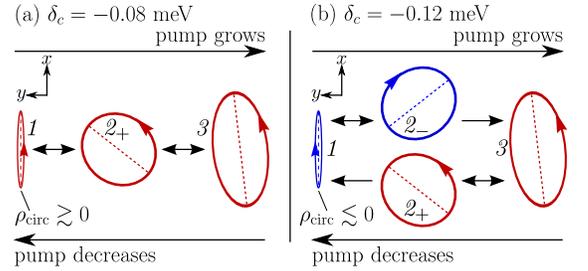}
  \caption{(Color online) The scheme of non-equilibrium transitions at
    (a) $\delta_c=-0.08$~meV and (b) $\delta_c=-0.12$~meV; in both
    cases $\delta_l = 0.05$~meV, $\rho_p = 0.5$, $\Phi_p=0$. Ellipses
    schematically represent the polarizations typical of the stability
    branches $\mathit 1$, $\mathit 2_\pm$, and $\mathit 3$ which are
    explicitly labeled in Fig.~\ref{fig:fig1}d-f.}
  \label{fig:fig2}
\end{figure}

In the simplest case of a degenerate system ($\delta_{c,l} = 0$) under
purely circular excitation ($\rho_p = 1$), the dependence of the
steady-state cavity-field intensity on pump intensity takes the form
of an `S'-shaped circuit~\cite{Baas04-pra,Gippius04-epl}
(Fig.~\ref{fig:fig1}a).  This implies, first, the possibility of a
jump (``switch-up'') in the cavity transmission and, second, the
hysteresis of the response under slowly changing pump power.  As long
as $|V_2| \ll V_1$, the response of the system excited with
elliptically polarized light may be treated in terms of the two
S-circuits corresponding to (almost uncoupled) $\sigma^+$ and
$\sigma^-$ polarization components.  A growth of pump density involves
two successive switch-ups, in the dominant $\sigma^+$ or $\sigma^-$
condensate component and then in the minor
one~\cite{Gippius07,Shelykh08-prl} (Fig.~\ref{fig:fig1}b).

In a magnetic field the response diagram becomes qualitatively
different if $\rho_p$ and $\delta_c = E_0^+- E_0^-$ have opposite
signs, i.\,e.\ if the pump polarization is biased towards that of the
lower split-off polariton level.  In this case each of the
$\sigma^\pm$ modes has its own advantage over the other: whilst the
lower one is pumped more ``intensively'' (e.\,g.\ $|F_+| > |F_-|$),
the upper one is closer to the resonance with the pump field that is
blue-detuned from both.  As a result, for a certain pump there may
coexist three high-energy branches corresponding to spin-up and
spin-down components switched-up individually and in combination
(Fig.~\ref{fig:fig1}c).  Due to the same reason the circular
polarization in the lowermost state can be either positive (right) or
negative (left) depending on $\Delta$ and~$\delta_{c,l}$.

Fig.~\ref{fig:fig1}d represents the system with $\rho_p = +0.5$,
$\delta_c = E_0^+ - E_0^- = -0.12$~meV, and $\delta_l = E_0^{(x)} -
E_0^{(y)} = 0.05$~meV; the main axis of pump polarization is
$x$-directed, so that $\Phi_p = 0$.  The X-Y splitting involves the
mixing of the $\sigma^\pm$ polarization components, which is sensitive
to the phase difference of $\psi_\pm$~\cite{Paraiso10,Bozat12}.  In
the considered case it extends the range of $I_p$ where only one of
the $\sigma^\pm$ components is found in the ``up'' position.  If the
sign of $\delta_l$ were negative (\emph{or} the pump polarization were
$y$-directed), the region with the two highly polarized stability
branches would be reduced~\cite{Gavrilov13-apl}.

\textit{Transitions between stability branches.} In a degenerate
system the order of the switch-ups is rigidly determined by the bias
of pump polarization; the reverse transitions observed in the course
of decreasing pump always proceed in the ``last-up, first-down'' order
(Fig.~\ref{fig:fig1}b).  In a magnetic field, such a behavior takes
place at small $\delta_c$ (Fig.~\ref{fig:fig2}a), but at larger
$\delta_c$ a new evolution scenario appears (Fig.~\ref{fig:fig2}b).
Given the system starts from the lower stability branch ($\mathit 1$)
in Fig.~\ref{fig:fig1}d, a slowly increasing pump successively
transfers it to branch $\mathit 2_-$ (that is mainly $\sigma^-$
polarized) and then to branch $\mathit 3$~\footnote{A direct $\mathit
  2_- \rightarrow 2_+$ transition, which might seem possible according
  to Fig.~\ref{fig:fig1}d, does not happen unless branch $\mathit 2_+$
  became the single available branch in a finite range of $I_p$; this
  would be the case for sufficiently larger $\delta_l$ values than
  that considered in our work}. However, the reverse transition
$\mathit 3 \rightarrow \mathit 1$ proceeds through the
$\sigma^+$-polarized branch $\mathit 2_+$.  Thus, the transitions
occur in the ``last-up, last-down'' order that allows the cavity-field
polarization to be inverted. It constitutes a qualitatively distinct
type of multistability from that in spin-degenerate polariton systems
\cite{Gippius07,Shelykh08-prl,Gavrilov10-en,Paraiso10,Sarkar10,Adrados10}.

As shown by numerical simulations performed for $70$~ps long pump
pulses (see below), the system undergoes either $\mathit 1 \rightarrow
\mathit 2_+$ or $\mathit 1 \rightarrow \mathit 2_-$ transformation
depending on the sign of polarization ($\rho_\mathrm{circ}$) in the
low-energy state at branch $\mathit 1$ (Fig.~\ref{fig:fig2}).  In
other words, a small externally controlled right or left bias of
\emph{intra-cavity} rather than external (pump) polarization
determines the branch the system comes to on reaching the threshold
pump power.

\begin{figure}
  \centering
  \includegraphics[width=0.905\linewidth]{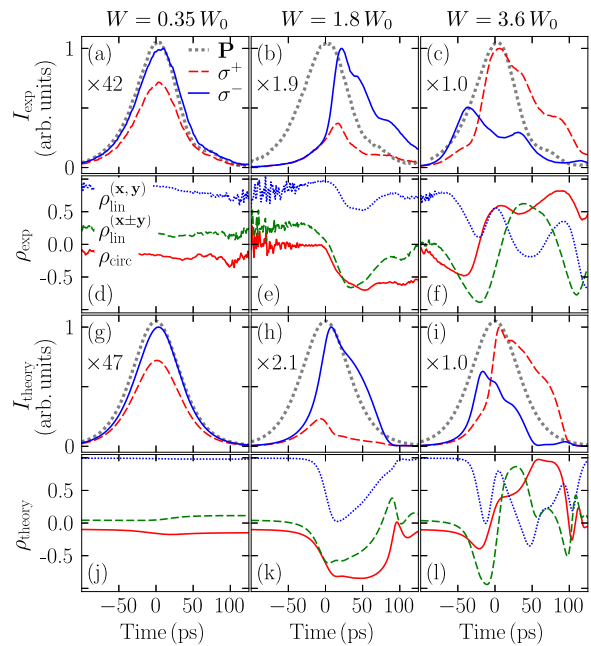}
  \caption{(Color online) (a--c) Measured time dependences of the
    $\sigma^+$ (solid lines) and $\sigma^-$ (dashed lines) components
    of the transmission signal at $\delta_c = -0.12$~meV; the pump
    shape is shown by heavy dotted lines. (d--f) The corresponding
    time dependences of the signal polarization in the circular basis
    (solid lines), $(\mathbf x, \mathbf y)$ basis (dotted lines), and
    $(\mathbf x \pm \mathbf y)$ basis (dashed lines).  Different
    columns correspond to different peak pump powers indicated at the
    top panels. (g--l) The modeled counterpart time dependences.  }
  \label{fig:fig3}
\end{figure}
\begin{figure}
  \centering
  \includegraphics[width=0.83\linewidth]{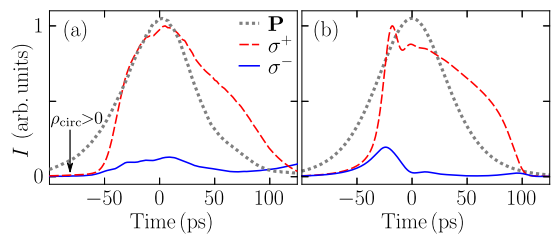}
  \caption{(a) Measured and (b) calculated time dependences of the
    $\sigma^+$ and $\sigma^-$ components of the transmission signal at
    $\delta_c = -0.08$~meV and $W = 3.6 W_0$.}
  \label{fig:fig4}
\end{figure}

\textit{Experiment.}---The sample has four 7~nm thick GaAs quantum
wells separated by 4~nm AlAs barriers which are centered in a
half-$\lambda$ cavity.  Its top (bottom) mirror consists of 32 (36)
Al$_{0.2}$Ga$_{0.8}$As/AlAs Bragg reflectors; GaAs substrate was
etched in order to perform transmission measurements.  The $Q$-factor
is $7 \cdot 10^3$; the Rabi splitting and the exciton-photon detuning
are 10.5~meV and $E_C - E_X(\mathbf k\,{=}\,0) \approx -5$~meV,
respectively; the decay rate of the lower polariton state at $\mathbf
k = 0$ is $\gamma_0 \approx 0.05$~meV.  The sample, placed into the
magneto-optical cryostat at $T = 2$~K, is excited by optic pulses with
a repetition rate of 8~MHz and duration of 70~ps generated by a
mode-locked Ti:sapphire laser. The pump beam is directed along the
cavity normal and focused into a $30~\mu$m wide spot on the sample.
The transmission signal is detected by a streak-camera with spatial
and time resolutions of $7~\mu$m and 6~ps, respectively.  Under a weak
excitation in a zero magnetic field the cavity spectrum exhibits two
linearly polarized levels separated by $\delta_l \approx 0.05$~meV.

To perform simulations, we solve numerically the many-mode
Gross-Pitaevskii equations of type (\ref{eq:gpe})~\cite{Gavrilov10-en}
with a $30~\mu$m wide Gaussian pump source; the model parameters
nearly coincide with the experimental ones.  In order to avoid
complexity brought on by spatial inhomogeneity of the transmission
(see \cite{Shelykh08-prl,Sarkar10,Adrados10}), below we only consider
the time dependences of the signal collected from the spot center (in
both the experiment and modeling).  At last, in order to simulate
finite time resolution and a partial weakly controlled time and space
smoothing of the measured signal, the calculated intensities are
averaged over the neighboring 10\,ps in each time point.

Fig.~\ref{fig:fig3} shows the dynamics of the system whose
steady-state properties have been discussed in
Figs.~\ref{fig:fig1}d--f.  Most importantly, the pump polarization is
biased towards the lower split-off circular polarization component
($\rho_p = +0.5$) and the upper linearly polarized eigenstate of a
spin-degenerate system (the $x$ direction).  The Zeeman splitting
$\delta_c \approx -0.12$~meV is involved by the magnetic field $B
\approx 6$~T.  The three basis polarization degrees of the signal, the
circular one, $\rho_\mathrm{circ}^{\vphantom\mathstrut}$, and two
linear ones, $\rho_\mathrm{lin}^{(\mathbf x, \mathbf y)}$ and
$\rho_\mathrm{lin}^{(\mathbf x \pm \mathbf y)}$, are shown in
Figs.~\ref{fig:fig3}d--f, whereas Figs.~\ref{fig:fig3}a--c show only
the circularly polarized components of the transmission
intensity. Left column in Fig.~\ref{fig:fig3} represents the case of a
low peak pump intensity $W = \max_t I_p(t) \approx 0.35 W_0$, where
$W_0$ corresponds to the first switch-up point in both the experiment
and calculation; in the experiment, $W_0 \approx
500~\mathrm{kW/cm^2}$.  In this case the signal polarization is nearly
constant, though it does not coincide with that of the pump beam due
to the splitting of polariton modes with different polarizations. In
fact, the signal is almost linearly polarized with a slight bias
towards $\sigma^-$ according to Figs.~\ref{fig:fig1}d--f.

The middle column in Fig.~\ref{fig:fig3} represents the
above-threshold dynamics at $W = 1.8\,W_0$; a 5-fold increase in $W$
involves more than a 20-fold increase in the transmission signal
(compare Figs.~\ref{fig:fig3}a,b). The signal appears to be mainly
$\sigma^-$-polarized, and nearby the peak excitation it also exhibits
a prominent polarization in the ``diagonal'' ($\mathbf x - \mathbf y$)
direction (see Fig.~\ref{fig:fig3}e). This makes up a clear
manifestation of the $\mathit 1 \rightarrow \mathit 2_-$ transition
(Fig.~\ref{fig:fig1}f).

A further two-fold increase in $W$ up to $3.6\,W_0$ (right column in
Fig.~\ref{fig:fig3}) allows us to track the double $\mathit 1
\rightarrow \mathit 2_- \rightarrow \mathit 3$ transition proceeding
during the growth of pump density as well as the ``asymmetric''
reverse transition $\mathit 3 \rightarrow \mathit 2_+$ at the back
front of the pulse, which are schematically shown in
Fig.~\ref{fig:fig2}b. In particular, according to Fig.~\ref{fig:fig3}c
the signal exhibits first the switch-up in the $\sigma^-$ component
(leading to high negative values of $\rho_\mathrm{circ}$) and next,
with increasing pump, the switch-up in $\sigma^+$ (so that
$\rho_\mathrm{circ}$ becomes positive).  Finally, the switch-down in
$\sigma^-$ at the back front results in a further growth of
$\rho_\mathrm{circ}$ up to values exceeding the pump polarization.
The transitions observed in all the three basis polarizations agree
with the simulation shown in Fig.~\ref{fig:fig3}i,l and conform to the
steady-state diagram in Fig.~\ref{fig:fig1}d--f.  According to the
latter, (i) $\rho_\mathrm{lin}^{(\mathbf x \pm \mathbf y)}(t)$
exhibits the sign reversal during the $\mathit 2_- \rightarrow \mathit
3$ transition and (ii) $\rho_\mathrm{lin}^{(\mathbf x, \mathbf y)}(t)$
has two minima, of which the first is immediately before the $\mathit
2_- \rightarrow \mathit 3$ transition ($t \approx -20$~ps) and the
second is immediately after the $\mathit 3 \rightarrow \mathit 2_+$
transition ($t \approx +50$~ps).

The discussed transitions differ from those in a system with small
$\delta_c$ splitting.  A typical dynamics in a 1.5 times smaller
magnetic field ($B = 4$~T, so that $\delta_c = -0.08$~meV, and $W =
3.6 W_0$) is presented in Fig.~\ref{fig:fig4}.  Its main distinction
is a positive sign of the signal polarization in a low-energy state
($\rho_\mathrm{circ} > 0$ at $t < -60$~ps), which predetermines the
$\mathit 1 \rightarrow \mathit 2_+$ trajectory under increasing pump
power.  A further increase in $W$ (not shown) involves the $\mathit
2_+ \rightarrow \mathit 3$ transition.  In fact, this scenario,
representing a steady-state scheme in Fig.~\ref{fig:fig2}a, is
qualitatively the same as that predicted~\cite{Gippius07} and
observed~\cite{Sarkar10} for a spin-degenerate system.  However, our
system is not affected by the long-lived excitonic reservoir (see
Refs.~\cite{Gavrilov12-prb,Sarkar10,Gavrilov10-jetpl-en,Vishnevsky12})
due to a comparatively short duration of pump pulses.


To summarize, we have discovered a mechanism of ultra-fast spin flips
in cavity-polariton condensates, which can be implemented in Bose
systems with energy-split eigenstates of opposite spins.  It is
involved by amplitude dependent shifts of eigenstates due to
spin-dependent inter-particle interactions.  This effect can be used
in the future highly tunable optic switchers and logic elements
working in the picosecond range or for creating short light pulses
with rapidly changing polarization.

The authors are grateful to V.\,B.\,Timofeev and S.\,G.\,Tikhodeev for
fruitful discussions. This work was supported by the RF President
grant MK-6863.2012.2, RFBR, and the State of Bavaria.


\end{document}